\documentclass[5p, sort&compress]{elsarticle}
\usepackage{graphicx}
\usepackage{multirow}
\usepackage{amsmath}
\usepackage{amssymb}
\usepackage{xspace}
\usepackage{gensymb}
\usepackage{hhline}
\usepackage{multirow}
\usepackage{booktabs}
\usepackage{hyperref}
\usepackage{cleveref}
\usepackage{aas_macros}
\usepackage{soul}
\usepackage{xcolor}
\usepackage{array}
\usepackage{adjustbox}
\usepackage{cancel}
\usepackage{fancyhdr}

\usepackage{bm}

\usepackage[normalem]{ulem}

\newcommand{\flash}{{\tt FLASH}\xspace}
\newcommand{\emu}{{\tt EMU}\xspace}

\newcommand{\varrhobar}{\ensuremath{\overline{\varrho}}\xspace}
\newcommand{\tmax}{\ensuremath{t_{\max}}\xspace}

\begin{document}

\title{Neutrino Fast Flavor Instability in three dimensions for a Neutron Star Merger}

\author[1]{Evan Grohs\corref{cor1}}
\ead{ebgrohs@ncsu.edu}
\cortext[cor1]{Corresponding author}

\author[2]{Sherwood Richers\corref{cor2}}
\cortext[cor2]{NSF Astronomy \& Astrophysics Postdoctoral Fellow}
\ead{richers@utk.edu}

\author[3,4,5]{Sean M. Couch}
\ead{scouch@msu.edu}

\author[6]{Francois Foucart}
\ead{francois.foucart@unh.edu}

\author[1]{James P. Kneller}
\ead{jim_kneller@ncsu.edu}

\author[1]{G. C. McLaughlin}
\ead{gcmclaug@ncsu.edu}

\affiliation[1]{Department of Physics, North Carolina State University, Raleigh, North Carolina 27695, USA}
\affiliation[2]{Department of Physics, University of California Berkeley, Berkeley, California 94720}
\affiliation[3]{Department of Physics and
Astronomy, Michigan State University,
East Lansing, Michigan 48824, USA}
\affiliation[4]{Department of Computational Mathematics, Science, and
Engineering, Michigan State University,
East Lansing, Michigan 48824, USA}
\affiliation[5]{Facility for Rare Isotope Beams,
Michigan State University,
East Lansing, Michigan 48824, USA}
\affiliation[6]{Department of Physics \&
Astronomy, University of New Hampshire,
Durham, New Hampshire 03824, USA}

\begin{abstract}

The flavor evolution of neutrinos in core collapse supernovae and neutron star mergers is a critically important unsolved problem in astrophysics.  Following the electron flavor evolution of the neutrino system is essential for calculating the thermodynamics of compact objects as well as the chemical elements they produce.
Accurately accounting for flavor transformation in these environments is challenging for a number of reasons, including the large number of neutrinos involved, the small spatial scale of the oscillation, and the nonlinearity of the system. 
We take a step in addressing these issues by presenting a method which describes the neutrino fields in terms of angular moments.
We apply our moment method to neutron star merger conditions and show it simulates fast flavor neutrino transformation in a region where this phenomenon is expected to occur.
By comparing with particle-in-cell calculations we show that the moment method is able to capture the three phases of growth, saturation, and decoherence, and correctly predicts the lengthscale of the fastest growing fluctuations in the neutrino field.  
\end{abstract}


\maketitle

\section{Introduction}

Neutron Star Mergers (NSMs) are dramatic multi messenger events producing a variety of observable signals. The first direct evidence for the synthesis of the heaviest elements --- the $r$-process or rapid-neutron-capture-process elements --- in neutron star mergers came from the optical/infrared emission (kilonova) associated with the first neutron star merger observed through gravitational waves, GW170817~\cite{2017PhRvL.119p1101A,2017ApJ...848L..13A,GBM:2017lvd,2017Sci...358.1559K,Cowperthwaite:2017dyu,2017Sci...358.1556C,2017ApJ...848L..32M,2017Natur.551...75S,2017ApJ...848L..16S,2017ApJ...848L..27T}.
This kilonova light curve decayed on a relatively slow timescale, requiring the presence of lanthanides in the ejected matter~\cite{2013ApJ...775...18B,Kasen:2013xka}. This confirmed many decades of prediction that neutron star mergers would make $r$-process elements \cite{1974ApJ...192L.145L,1976ApJ...210..549L}.  While the original proposal for $r$-process in NSMs required tidal stripping, there are now several additional types of ejecta that are proposed to make $r$-process elements in NSM, including the collisional ejecta~\cite{2013PhRvD..87b4001H,2018ApJ...869..130R}, viscous evaporation~\cite{2013MNRAS.435..502F} and magnetically-~\cite{Siegel:2017nub,Fernandez:2018kax} or neutrino-driven winds~\cite{Surman:2005kf,Perego2014}.  As the neutrinos affect not only the thermal evolution of the neutron star merger, but also the number of neutrons available for capture in the ejecta~\cite{Surman:2003qt,Wanajo2014} and therefore the shape and magnitude of the kilonova signal, e.g., \cite{Zhu:2018oay,Barnes:2020nfi}, the investigation of the neutrino dynamics is of paramount importance.
As neutrinos propagate through the media, their numbers, energies, momenta, and flavors will change as functions of time and position.
In particular, an accurate accounting of the flavor of the neutrinos as a function of position and time is essential.

In some large scale dynamical simulations of NSMs neutrinos are already included, albeit classically, as they facilitate energy transfer and charge exchange in the evolving compact object.
Much work has been done to characterize the transport processes which change number, energy, and momenta.  
There exists an extensive literature on using different techniques to capture neutrino transport in compact objects, taking inspiration from methods developed for photon and neutron transport.
Moment methods \cite{thorne80,10.1143/PTP.125.1255, 10.1093/mnras/194.2.439, PhysRevD.87.103004, 10.1093/mnras/stac589,FoucartM1:2016,Radice:2021jtw} approximately reproduce the behavior of neutrino transport in many cases, and their efficiency enables simulations of a much larger scale than would be possible with direct Boltzmann transport (e.g., \cite{10.1093/mnras/stz3223,2018ApJ...854...63O,Radice:2021jtw, Just:2022flt,roberts_GeneralrelativisticThreedimensionalMultigroup_2016,Li:2021vqj}).  However, due to their approximate nature, moment methods always require validation against more exact grid-based \cite{Liebendorfer_2005,PhysRevD.63.103004,Liebendorfer_2005,Mezzacappa:2020oyq,O_Connor_2018,10.1093/mnras/stab2983} or Monte Carlo \cite{1992A&A...256..452J,Richers_2015,PhysRevD.98.063007,Miller:2019dpt,Foucart2020_weUsedThe2018PaperNotThisOne,Kato_2020,kawaguchi_MonteCarloBasedRelativistic_2022,Foucart:2020qjb} Boltzmann transport calculations.
In this manuscript, we adapt an existing two-moment neutrino transport scheme to  simulate the transport of neutrino flavor.

Not only do neutrinos change flavor due to the non-zero neutrino rest mass, the oscillations are modified by coherent forward scattering on neutrons, protons and electrons, i.e., the MSW effect \cite{Mikheyev:1985aa,Mikheyev:1986tj,1978PhRvD..17.2369W} which operates on the outskirts of an NSM. In addition, deep in the interior of an NSM, coherent forward scattering on other neutrinos, 
the so-called neutrino self interaction effects, can play a dominant role \cite{Zhu:2016mwa,Frensel:2016fge,Tian:2017xbr}. The transformations due to the self interaction can be roughly grouped into three classes: the matter-neutrino resonance \cite{Malkus:2012ts,2018PhRvD..97h3011V}, collective oscillations \cite{2006PhRvD..74l3004D} and Fast Flavor Oscillations (FFO) \cite{Sawyer:2005jk}.  
The importance of fast flavor transformation in mergers has been previously pointed out using parameterized outcomes of the FFO, e.g. \cite{PhysRevD.84.053013,Wu:2017drk,Li:2021vqj,Just:2022flt}.
Multi-angle methods have been used to simulate the FFO directly in idealized scenarios or localized domains (e.g., \cite{duan_FlavorIsospinWaves_2021,tamborra_NewDevelopmentsFlavor_2021,george_COSENuCollective_2022,bhattacharyya_ElaboratingUltimateFate_2022,richers_CodeComparisonFast_2022,nagakura_TimedependentQuasisteadyGlobal_2022,richers_FastFlavorTransformations_2022}).  
However multi-angle methods are computationally expensive and the expense is a barrier to studying FFO in more realistic geometries. Using moments would significantly reduce the computational expense.
But FFO are sensitive to the angular distributions -- something the moments integrate over -- so it is not clear whether moment methods are capable of capturing FFO.

In this work, we present the first simulations of quantum moment based neutrino FFO using neutrino distributions taken directly from a general-relativistic simulation of an NSM. 
We find that for these conditions, a moment based approach is capable of qualitatively capturing the local phenomenon and to determine its accuracy, we compare results with particle-in-cell calculations.

\section{Evolution Equations and Methods}

Our starting point is the result of the classical global general relativistic two-moment radiation hydrodynamics simulation of Ref.\ \cite{Foucart:2016rxm}
which provides moments of the neutrino distributions for an NSM remnant. 
Quantum coherence of multi-neutrino wave functions require us to treat the system as an ensemble of $n$ particles with a corresponding density matrix.
If we ignore higher-order correlations by employing a mean-field approximation, we can construct dynamical quantities which only encode flavor coherence. (For investigations of the appropriateness of the mean field approximation see \cite{2003PhRvD..68a3007F,2003JHEP...10..043F,Rrapaj:2019pxz,Patwardhan:2021rej,Lacroix:2022krq,Roggero:2022hpy}). These quantities are the neutrino generalized density matrices, $\varrho$ and \varrhobar which characterize the phase-space density for neutrinos and antineutrinos at time $t$, 3-position $\mathbf{x}$, and 3-momentum $\mathbf{p}$. When integrated over phase space, their diagonal components represent the number of neutrinos of a given flavor while the off-diagonal components give a measure of the coherence between two flavors.

References \cite{1993NuPhB.406..423S,Volpe:2013uxl,2014PhRvD..89j5004V,2016PhRvD..94c3009B} derive evolution equations for $\varrho$ and \varrhobar which are commonly known as the Quantum Kinetic Equations (QKEs) and involve a Hamiltonian-like operator, $H$
\begin{align}
  \frac{\partial\varrho}{\partial t} + \dot{\mathbf{x}}\cdot\frac{\partial\varrho}{\partial\mathbf{x}} + \dot{\mathbf{p}}\cdot\frac{\partial\varrho}{\partial\mathbf{p}}
  &= -i[H,\varrho] + C,\label{eq:qke_nu}\\
  \frac{\partial\varrhobar}{\partial t} + \dot{\mathbf{x}}\cdot\frac{\partial\varrhobar}{\partial\mathbf{x}} + \dot{\mathbf{p}}\cdot\frac{\partial\varrhobar}{\partial\mathbf{p}}
  &= -i[\overline{H},\varrhobar] + \overline{C},\label{eq:qke_bnu}
\end{align}
where the dot over a quantity represents differentiation with respect to time.  The left-hand-side (lhs) of Eqs.\ \eqref{eq:qke_nu} and \eqref{eq:qke_bnu} give the evolution of the density matrices through phase space.  We will set $\dot{\mathbf{p}}=0$ when we consider flat Minkowski space for our calculations.
The rhs of Eqs.\ \eqref{eq:qke_nu} and \eqref{eq:qke_bnu} gives a generic source term which governs the evolution of the density matrices.  $C$ and $\overline{C}$ are momentum-changing collision terms which we will neglect here.  The Hamiltonian-like operators are sums of three terms: a vacuum potential which mixes flavor states during neutrino propagation; a matter potential which gives rise to forward scattering of neutrinos on background electrons, positrons, and nucleons; and a self-interacting potential $H_\nu$ due to forward scattering on the other neutrinos, namely
\begin{equation}
  H_{\nu} = \frac{\sqrt{2}\,G_F}{(2\pi)^3}\int d^3q\left(1-\frac{p_j\,q^j}{p\,q}\right)[\varrho(t,\mathbf{x},\mathbf{q})-\varrhobar^{\ast}(t,\mathbf{x},\mathbf{q})],
  \label{eq:SI1}
\end{equation}
where $G_F=1.166\times10^{-11}\,{\rm MeV}^{-2}$ is the Fermi constant, $\mathbf{p}$ is a free variable for the density matrix in the commutators of Eqs.\ \eqref{eq:qke_nu} and \eqref{eq:qke_bnu}, and $\mathbf{q}$ is a dummy variable of integration.

The QKEs may be recast as a tower of evolution equations for the angular moments of the density matrices \cite{2013PhRvD..88j5009Z,2018PhRvD..98j3001D,Johns:2019izj,2020PhRvD.102j3017J}. The first three angular moments are more familiarly known as the energy density $E$, the flux vector ${\bf F}$ and the pressure tensor $P$. The explicit expressions for these moments are
\begin{align}
  E_{ab}(t,\mathbf{x},p) &= \frac{p^3}{(2\pi)^3}\int d\Omega_p\, \varrho_{ab}(t,\mathbf{x},\mathbf{p}),\label{eq:mom_0}\\
  F^j_{ab}(t,\mathbf{x},p) &= \frac{p^3}{(2\pi)^3}\int d\Omega_p\frac{p^j}{p}\, \varrho_{ab}(t,\mathbf{x},\mathbf{p}),\label{eq:mom_1}\\
  P^{jk}_{ab}(t,\mathbf{x},p) &= \frac{p^3}{(2\pi)^3}\int d\Omega_p\frac{p^jp^k}{p^2}\, \varrho_{ab}(t,\mathbf{x},\mathbf{p}),
\label{eq:mom_2}
\end{align}
where $p=|\mathbf{p}|$, subscripts $a,b\in\{e,\mu,\tau\}$ represent flavor indices, and superscripts $j,k\in\{x,y,z\}$ represent space indices. The factor of $4\pi$ in these definition is a normalization convention. 
As the phenomenon of fast flavor transformation is driven by the self-interaction, it is sufficient to write the first two moment evolution equations as 
\begin{align}
  \frac{\partial E}{\partial t} + \frac{\partial F^j}{\partial x^j}
  &= -i[H_E,E] + i[H_{F}^{j},F_j],\label{eq:qke_E}\\
  \frac{\partial F^j}{\partial t} + \frac{\partial P^{jk}}{\partial x^k}
  &= -i[H_E,F^j] + i[H_{F}^{k},P^{j}_{k}],\label{eq:qke_F}
\end{align}
keeping only the self-interacting term of the Hamiltonian while neglecting the matter and vacuum contributions. These equations have implicit summations over repeated indices $j$ and $k$ using a Euclidean metric, and where
\begin{align}
  H_E &= \sqrt{2}\,G_F\int_0^{\infty} \frac{dq}{q}\left(E-\overline{E}^{\star}\right),\label{eq:he}\\
  H_{F}^{j} &= \sqrt{2}\,G_F\int_0^{\infty} \frac{dq}{q}\left(F^j-\overline{F}^{j\star}\right).\label{eq:hf}
\end{align}
The partial derivatives with respect to space denote advection and are key to capturing inhomogeneous modes of FFO.
Similar equations can be derived for the evolution of the antineutrino energy density, $\overline{E}$ and flux vector $\overline{F}^j$.

The recasting of the QKEs as evolution equations for the moments is not an approximation: the moment evolution equations are simply a series of linear combinations of the QKEs with weighting factors proportional to the components of the neutrino momentum unit vectors. However, truncating the series of moment equations  does involve an approximation since a priori we do not know the exact closure -- the relationship between the lowest order unevolved moment [$P^{jk}$ in Eq.\ \eqref{eq:qke_F}] and the evolved moments.
We algebraically solve for $P^{jk}$ by stipulating a specific closure relationship: the Maximum Entropy Closure (MEC).  The MEC relies on an interpolation between the optically thin and thick limits (or more properly, neutrino thin and thick limits)
\begin{equation}
  P_{ab}^{jk} = E_{ab}\left[ \left(\frac{3 \chi -1}{2}\right) \frac{F_{ab}^j F_{ab}^k}{|{\bf F}_{ab}|^2} +
  \left(\frac{1- \chi}{2} \right) \delta^{jk}\right],
  \label{eq:closure}
\end{equation}
where $|{\bf F}_{ab}|$ is the magnitude of the flux vector for the $ab$ component of $\varrho$ and $\delta^{jk}$ is the Kronecker delta function.  Analogous expressions exist for anti-neutrino moments $\overline{E},\overline{F}^j,\overline{P}^{jk}$.
The quantity $\chi$ ($\overline{\chi}$) is analogous to the classical Eddington factor, and we use the flavor traced flux factor in determining this quantity for all density matrix components \cite{usinprep}.
We calculate $\chi$ based on the MEC relation.
(For more detail on the MEC in the classical limit, see \cite{1994ApJ...433..250C,2017MNRAS.469.1725M}.) 
For the flavor off-diagonal components of $P_{ab}^{jk}$, we evaluate Eq.\ \eqref{eq:closure} for the real part of $P_{ab}$ using the real part of $f_{ab}$, and separately calculate the imaginary part of $P_{ab}$ using the imaginary part of $f_{ab}$. Future improvements will be needed to remove the basis dependence that stems from this prescription.

Many codes already exist which contain the infrastructure for organizing moment variables
(including Refs.\ \cite{Skinner_2019, 10.1093/mnras/stac589, Laiu_2021, Bruenn_2020, Roberts_2016, 10.1093/mnras/stv1892, Kuroda_2016}, among many others) and time-evolving those variables by calculating the advection terms on the left hand side of Eqs.\ \eqref{eq:qke_E} and \eqref{eq:qke_F}.  We choose the \flash code \cite{2000ApJS..131..273F} and amend it to include neutrino flavor transformation.  
Modifications include adding in new variables for the off-diagonal terms of the energy density and flux moments, and adding in new subroutines to calculate the commutators in Eqs.\ \eqref{eq:qke_E} and \eqref{eq:qke_F}. Although we use energy density in our 
moment scheme, we only consider monoenergetic neutrinos in this work, implying the relationship between number and energy densities is simply
$N_{ab}(\mathbf{x},p) = (1/p) E_{ab}(\mathbf{x},p)$.
For a basis of comparison to the \flash results, we also do comparable QKE calculations with the Particle-In-Cell (PIC) quantum kinetics code \emu \cite{2021PhRvD.103h3013R,2021PhRvD.104j3023R}. \emu follows the evolution of neutrino quantum states along discrete directions instead of evolving angular moments.

\begin{table}
    \centering
    \begin{tabular}{c|c}
         Input & Value \\ \hline
         $N_{ee}\,({\rm cm}^{-3})$ & $1.422\times10^{33}$\\
         $\overline{N}_{ee}\,({\rm cm}^{-3})$ & $1.915\times10^{33}$\\
         $\Sigma N_{(x)}\,({\rm cm}^{-3})$ & $1.965\times10^{33}$\\
         $\mathbf{f}_{ee}$ & $(\phantom{-}0.0974, \phantom{-}0.0421, -0.1343)$\\
         $\mathbf{\overline{f}}_{ee}$ & $(\phantom{-}0.0723, \phantom{-}0.0313, -0.3446)$\\
         $\mathbf{f}_{(x)}$ & $(-0.0216, \phantom{-}0.0743, -0.5354)$\\
    \end{tabular}
    \caption{
    List of simulation parameters initially taken from Ref.\ \cite{Foucart:2016rxm} at the black cross in Fig.\ \ref{fig:foucart}. The first three rows show the number densities of each neutrino flavor. For clarity, the third row shows the sum of all four heavy lepton anti/neutrino densities. Three-flavor simulations assume $N_{\mu\mu}=\overline{N}_{\mu\mu}=N_{\tau\tau}=\overline{N}_{\tau\tau}=\Sigma N_{(x)}/4$, while two-flavor simulations assume $N_{xx}=\overline{N}_{xx}=\Sigma N_{(x)}/4$. The next three rows show the flux factor vectors, the norm of which are the flux factors. The last row is the flux factor vector for a heavy-lepton in either the two or three flavor simulations. 
    }
    \label{tab:parameters}
\end{table}

\begin{figure}
    \centering
    \includegraphics[width=\linewidth]{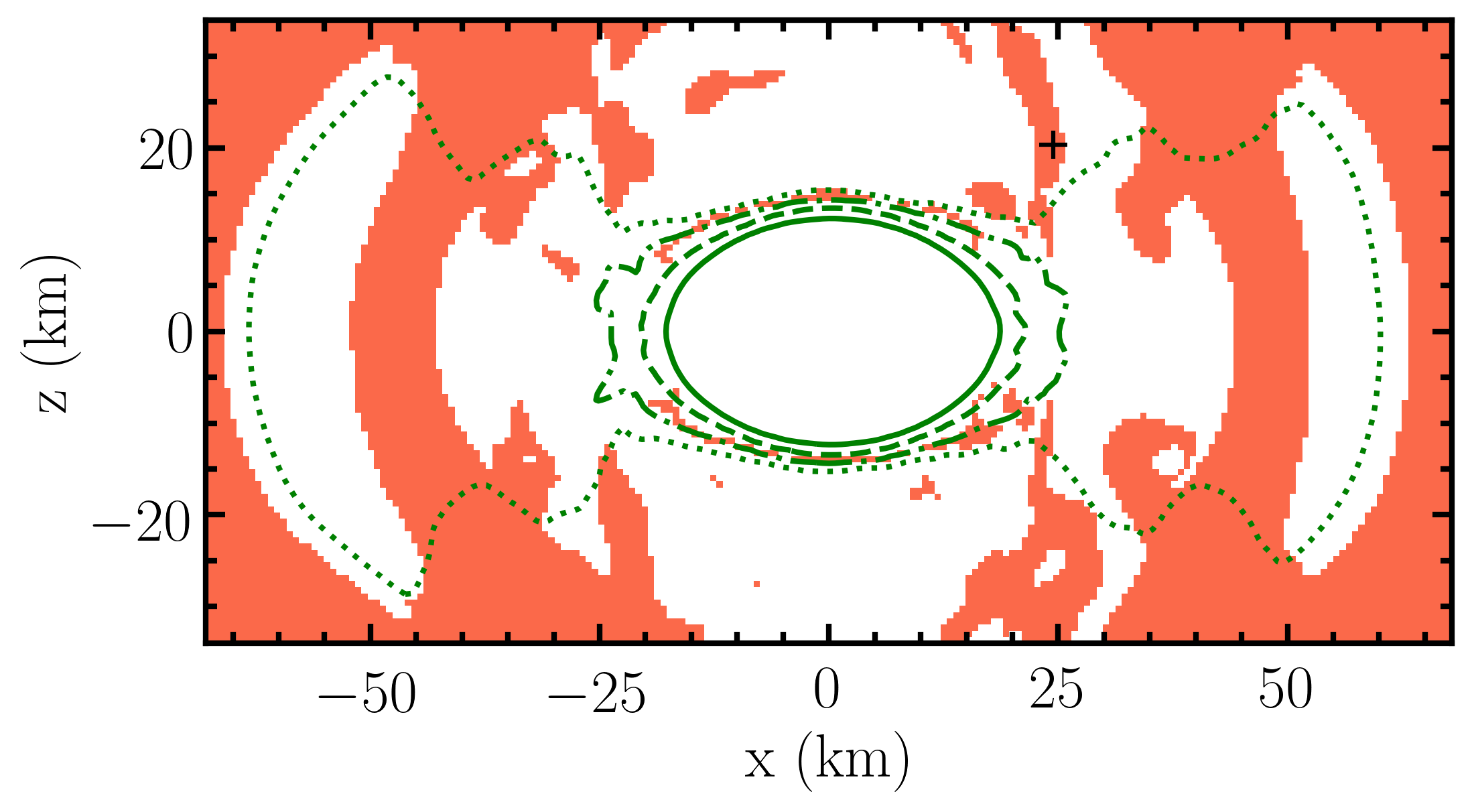}
    \caption{Locations of a Lepton-Number crossing in a neutron star merger simulation $5\,\mathrm{ms}$ after merger as indicated by the presence of a zero crossing in the electron lepton number current. The radiation field energy density and flux come from the simulation of \cite{Foucart:2016rxm}. Neutrino distributions are assumed to follow the angular distribution described by the maximum entropy closure, consistent with the original simulation.
    A colored region indicates that there exists at least one direction where the flow of electron-flavor neutrinos is equal to the flow of electron-flavor antineutrinos, equivalent to a zero crossing in electron lepton number as determined using the maximum entropy crossing test in \cite{2022arXiv220608444R}.  White regions indicate no such direction exists.
    The black cross indicates the location from which the parameters for calculation were drawn. From outside to inside, the green contours indicate matter densities of \{$10^{11},10^{12},10^{13},10^{14}$\} $\mathrm{g\,cm}^{-3}$.}
    \label{fig:foucart}
\end{figure}

A necessary condition for fast flavor transformation is that the 
lepton number flux
as a function of direction changes sign \cite{Capozzi:2020syn,Johns:2021taz,2021PhRvD.104h3035Z}.  We look for spatial points where these crossings occur using the values of the angular moments from the merger simulation \cite{Foucart:2016rxm}, by first transforming from curvilinear coordinates into a comoving orthonormal tetrad to allow us to take advantage of the local flatness of the metric, rotating the coordinates to align the net electron lepton number flux with the $z$-axis, and then applying the maximum entropy crossing test of Ref.\ \cite{2022arXiv220608444R}.  Figure \ref{fig:foucart} shows a polar slice of the merger simulation 5 ms after the collision, where shaded regions indicate instability. 
The black cross at $(x,z)\sim(25,20)\,{\rm km}$ shows the location from which we extract the initial conditions.  We deliberately choose a point close to the disk where the flavor content of the neutrino field is likely to influence the wind nucleosynthesis \cite{Surman:2005kf,Surman:2008qf,Malkus:2012ts}.
Table \ref{tab:parameters} gives the initial conditions for the neutrino moments at the black cross.

Emerging from the dense environment of the remnant, we would expect the neutrino worldlines to follow different paths from the site of the last scattering \cite{2012PhRvL.108z1104C,2013PhRvD..87h5037C}.
The vacuum term in the QKEs will act to populate the flavor-off-diagonal components with different phases for the different neutrino worldlines. 
That associated potential is smaller than the self-interacting potential of Eqs.\ \eqref{eq:he} and \eqref{eq:hf}, where we estimate the ratio to be $10^{-6}$ at the conditions present in Fig.\ \ref{fig:foucart}.
To incorporate phases into the off-diagonal components of our moment simulation, we perturb the imaginary and real components of the energy density moment by using random numbers scaled by the ratio $10^{-6}$ of the maximum value of the flavor-diagonal components
\begin{equation}\label{eq:od_seed}
  \delta E_{ab}(\mathbf{x}) 
  = 10^{-6}\,p\max\{N_{cc}\}[A_{ab}(\mathbf{x})+iB_{ab}(\mathbf{x})]
\end{equation}
where $-1<A,B<1$, and $a\ne b$.
For the flux moment, we implement the same perturbation as the energy density moment using flux factors
$\delta\mathbf{F}_{ab}(\mathbf{x}) = \delta E_{ab}(\mathbf{x})(\Sigma_c N_{cc}\mathbf{f}_{cc})/$ $(\Sigma_c N_{cc})$,
implying the off-diagonal components for $E$ and $\mathbf{F}$ are correlated.  Analogous expressions exists for the anti-neutrino moments.
For the PIC method, we give the density matrices $\varrho$ nonzero off-diagonal components with the same $10^{-6}$ scaling and randomization for {\it each} particle and each off-diagonal element in each cell.
If we calculate the energy density for the PIC method using Eq.\ \eqref{eq:mom_0}, we would expect the incoherent sum for $\delta E_{ab}$ to be reduced by $\sqrt{n}$ for $n$ particles per cell.
As a result, we expect different behavior between the moment and PIC methods due to a different initialization of $\delta E_{ab}$.
Note that we see identical behavior with smaller perturbations since the perturbations grow exponentially.

\section{Results}

To solve the QKEs in both the moment and PIC methods, we use a 3-dimensional Cartesian geometry.  The system is a cube with side length $L=7.87\,{\rm cm}$.  We divide the domain into 128 cells per side and institute periodic boundary conditions.
Our choice of geometry allows us to study a range of spatial modes and resolve the fastest growing one, with little interference from boundary effects of a finite domain.
For the flavor evolution, the moment method uses 2 flavors, $e$ and a heavy-lepton $x$.

\begin{figure}
    \centering
    \includegraphics[width=\linewidth]{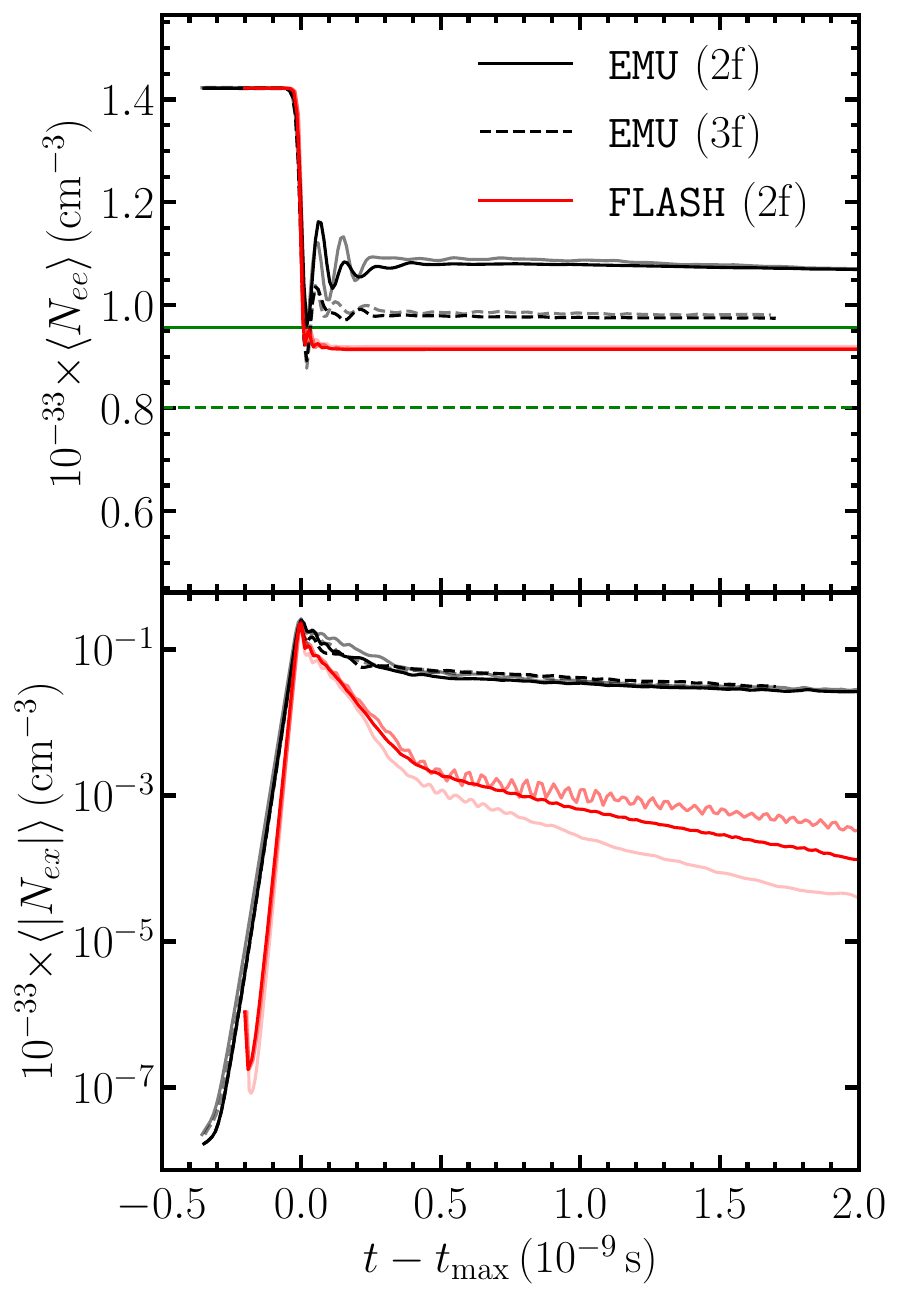}
    \caption{Spatially averaged components for the number density moment plotted against time measured from the point of saturation.  [Top] Diagonal $ee$ component plotted against $t-\tmax$.  Solid (dashed) black lines show the results for the 2-flavor (3-flavor) \emu baseline calculations.  The solid red line shows the results for the \flash baseline calculation.  Also plotted are the results for the simulations with a smaller domain but identical resolution as gray and medium red lines.
    In addition, we give a light red line for a calculation with a smaller domain but double the resolution of the baseline calculation.
    Finally, we give a solid (dashed) green line to show the value of $N_{ee}$ for 2-flavor (3-flavor) equilibration.
    [Bottom] Magnitude of off-diagonal $ex$ component plotted against $t-\tmax$.  Color and linestyle conventions same as top panel.  For the 3-flavor \emu calculation, $ex=e\mu$.  The growth rate for \emu (2f) is $5.58\times10^{10}\,{\rm s}^{-1}$ and for \flash is $8.09\times10^{10}\,{\rm s}^{-1}$.}
    \label{fig:P_NSM}
\end{figure}

Figure \ref{fig:P_NSM} gives averaged components of the neutrino number density moment $N_{ab}$ plotted against $t-\tmax$, where \tmax is the time at the point of saturation ($\tmax\simeq0.5\,{\rm ns}$ for \emu simulations; $\tmax\simeq0.2\,{\rm ns}$ for \flash).
The solid dark red curve shows the results from the \flash simulations, whereas the solid (dashed) black curve gives the results from the 2-flavor (3-flavor) \emu calculations.  To test for convergence, we also give results for other simulations where we changed the domain parameters.  
Instead of using a cube with side-length $7.87$ cm, the additional simulations have a side-length of 3.93 cm.  The number of grid points per side also differs.  The gray lines for both \emu calculations, and the medium red line for \flash have 64 grid points per side, yielding the same resolution as the baseline case.  The light red line for \flash has 128 grid points per side, implying double the resolution of the baseline case.  We observe that the \flash calculations of differing domain size and resolution begin to diverge for $t-\tmax\simeq0.4\,{\rm ns}$.  This is due to an accumulation of error in the finite differencing of the advective term in Eqs.\ \eqref{eq:qke_E} and \eqref{eq:qke_F}.  Therefore, we focus our discussion on times before $t-\tmax\simeq0.4\,{\rm ns}$, which includes the growth and saturation phases of the FFO.

We first examine the growth of the instability which can be seen as the abrupt drop of all curves in $N_{ee}$ at $t-\tmax = 0$ (top panel) and the corresponding rise in $|N_{ex}|$ (bottom panel).  $|N_{ex}|$ grows from the initial perturbation until saturation, where the slope of the line is the exponential growth rate. We see excellent qualitative agreement between the moment and PIC methods, with our \flash simulations having a slightly faster growth rate ($8.09\times10^{10}\,{\rm s}^{-1}$) than our \emu calculations ($5.58\times10^{10}\,{\rm s}^{-1}$ for 2-flavor).

Second, we examine the point of saturation, corresponding to the peak of $\langle|N_{ex}|\rangle$ in the bottom panel.  The magnitude of this peak is nearly identical 
for both methods.  \emu has a larger duration of time where $\langle|N_{ex}|\rangle$ is large, and we see this reflected in the top panel where the amplitude of the first oscillation in $\langle N_{ee}\rangle$ is larger for \emu than \flash, indicating more flavor transformation in the former. 

Third, we examine the decoherence phase, which corresponds to the leveling out of the $\langle N_{ee}\rangle$ curves in the top panel.
We see that the electron number density agrees to about 30\% for our choice of closure.
For reference, the green lines show the value of $\langle N_{ee}\rangle$ if the system had reached a 2-flavor equilibration (solid) or 3-flavor equilibration (dashed).  Examining the bottom panel, we see a larger loss of coherence in the \flash calculation as well as the resulting freeze-out of $\langle N_{ee} \rangle$ at a smaller value as compared to \emu.

\begin{figure}
    \centering
    \includegraphics[height=0.6\linewidth,trim={2cm 0cm 5cm 0cm},clip]{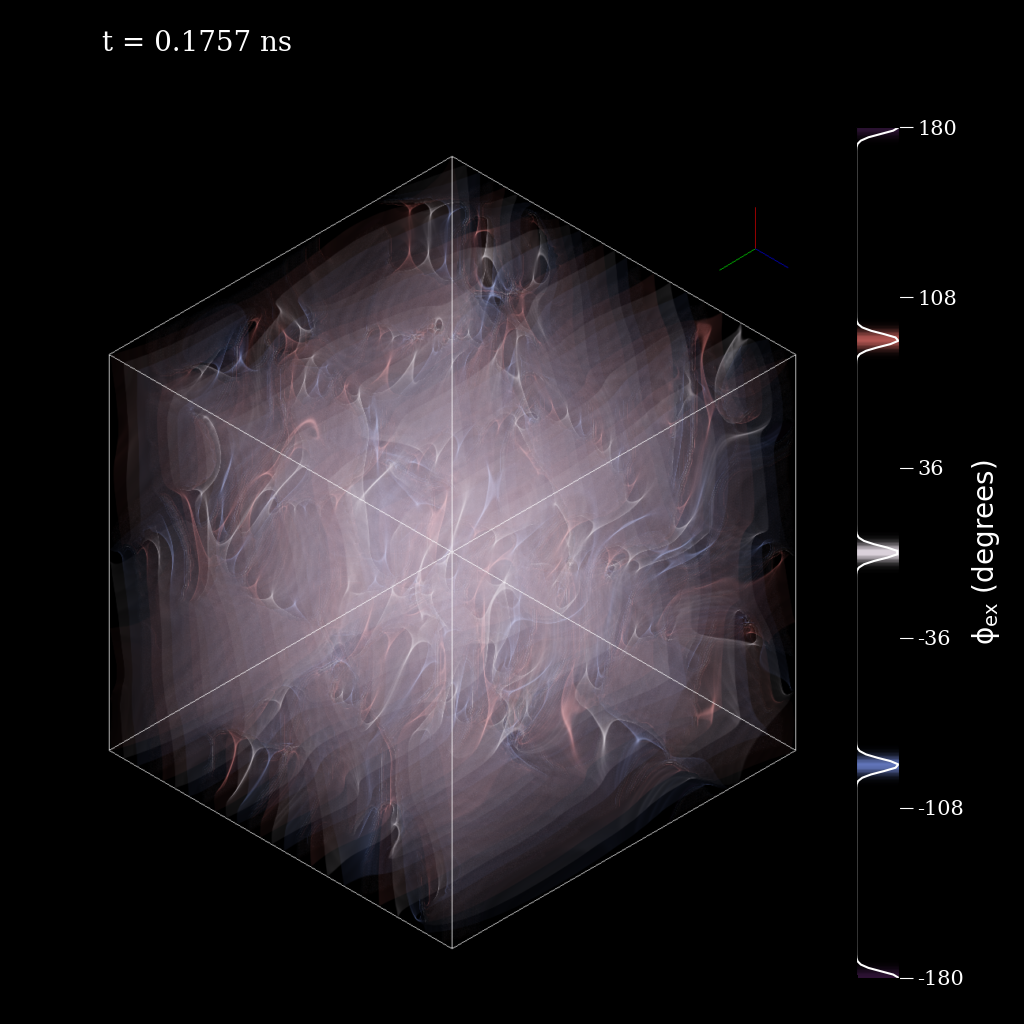} \hspace*{-0.9em}
    \includegraphics[height=0.6\linewidth,trim={2cm 0cm 0cm 0cm},clip]{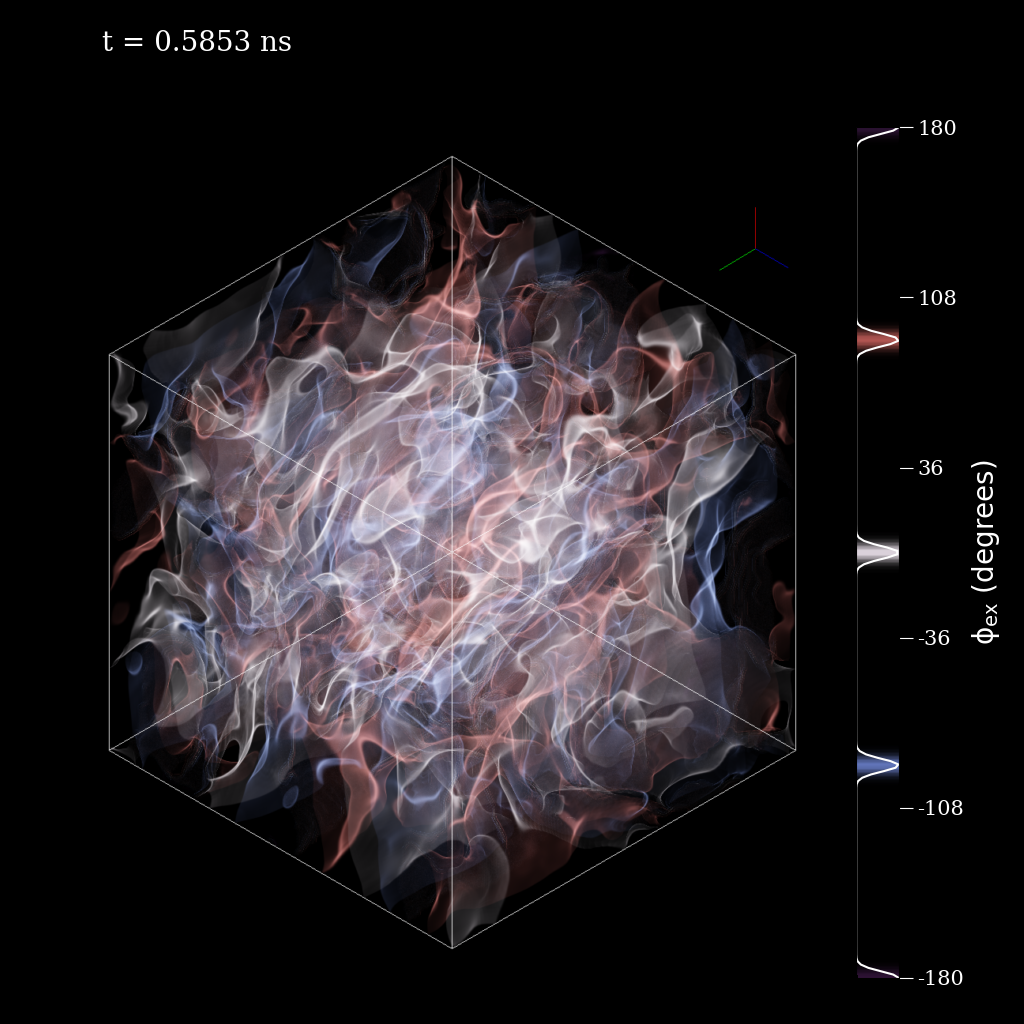}
    \caption{
    Volume rendering of contours of $\phi_{ex}$ (the phase of the number density moment $N_{ex}$) during the linear growth phase (left) and after saturation (right). The geometry is a cube with side length $L=7.87\,{\rm cm}$.  In the growth phase, approximately planar structures show wavefronts of the fastest growing mode. After saturation the phases chaotically mix.
    }
    \label{fig:volume_rendering}
\end{figure}

To show the behavior of the phase of the off-diagonal term $\phi_{ex}$, we plot 3D contours of the phase at 2 different time slices for the \flash calculation in Fig.\ \ref{fig:volume_rendering}.  The left panel shows the linear growth phase, where extended wavefronts of equal phase have begun to form. These waves grow larger in amplitude until the point of the FFO saturation.  Although there is a net decoherence after saturation, the characteristic wavelength of the phase variations increases (right panel).  This qualitative behavior is similar to what we see in PIC simulations, where we perform the calculation with both 2 and 3 flavors.

\begin{figure}
    \centering
    \includegraphics[width=\linewidth]{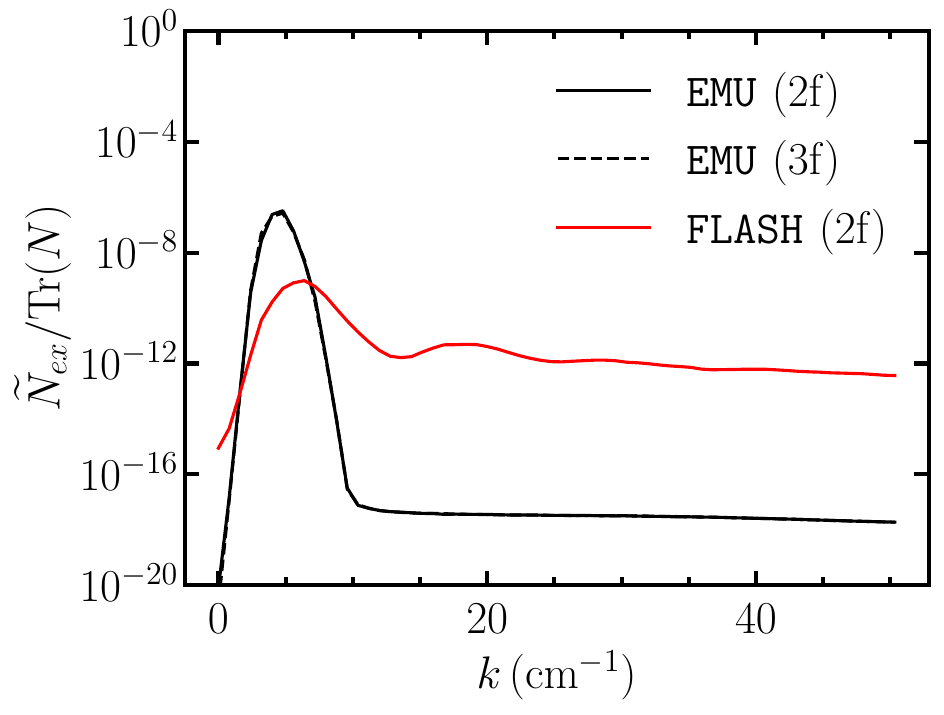}
    \caption{Discrete Fourier transform of $N_{ex}$ plotted against wavenumber $k$. $k$ is the magnitude of the 3-vector.  Linestyles and colors are the same as Fig.\ \ref{fig:P_NSM}.  The time is taken as $t-\tmax=-0.1\,{\rm ns}$ for all three curves.
    The mode with the peak power for \emu (2f) is $4.79\,{\rm cm}^{-1}$ and for \flash is $6.39\,{\rm cm}^{-1}$. The rightmost end of the curves shows power at the grid scale. Lower resolution simulations (not shown) truncate at smaller $k$, but have a peak in the same location.
    }
    \label{fig:FFT_NSM}
\end{figure}

Fast flavor transformation is a phenomenon where the rapid growth occurs only on certain fluctuation scales as illustrated in Fig.\ \ref{fig:volume_rendering}. 
To determine this scale 
we show a Discrete Fourier Transform of the complex $N_{ex}$ component normalized by the flavor trace of $N_{ab}$ versus the magnitude of the wavenumber in Fig.\ \ref{fig:FFT_NSM}.
Both the \emu and \flash calculations show power peaking at nearly the same wavevector $k\sim5\,{\rm cm}^{-1}$, corresponding to the fastest growing mode on a scale $\lambda=2\pi/k\sim1.0\,{\rm cm}$.  Thus the moment method effectively captures the fluctuation scale as well.  There are some small differences that would likely change if a different closure were applied in the moment method.  The peaks are slightly offset and there is a larger amount of power in the \emu calculations at the peak.  In addition, there is more power in the \flash calculations at larger wavenumber, which we expect to have a numerical origin and be intrinsic to the moment scheme.

\begin{figure*}
    \centering
    \includegraphics[width=\linewidth]{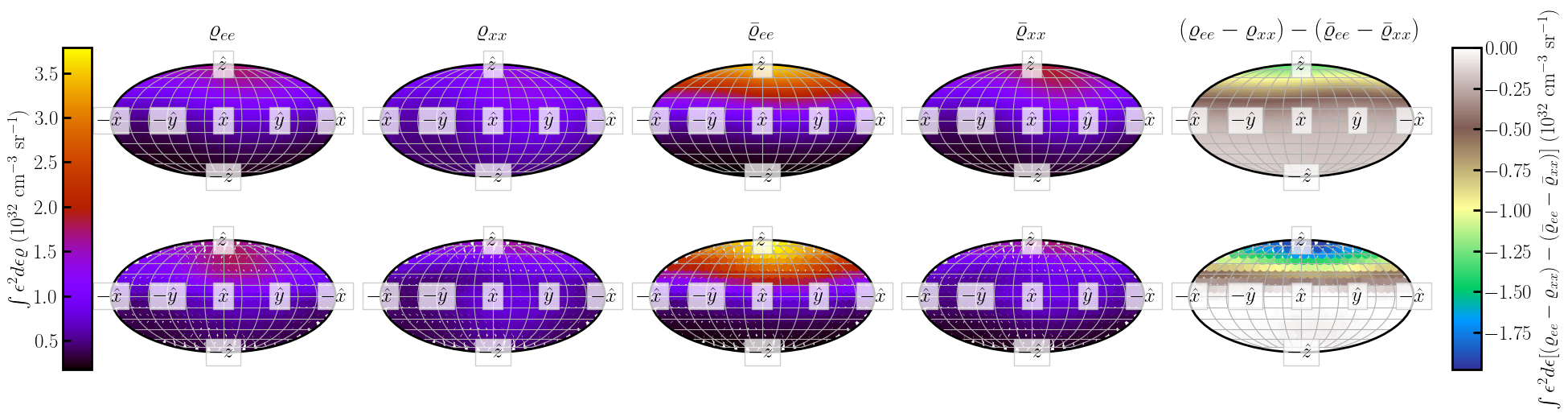}
    \caption{Energy-integrated and space-averaged angular distributions at $t=2\,\mathrm{ns}$. The top row shows the distributions implied by applying the MEC to the simulated moments from the \flash calculation, where each moment is averaged over the spatial domain. The bottom row shows the results from the \emu calculations, where each direction is averaged over the spatial domain and plotted as a colored dot.}
    \label{fig:angular_distributions}
\end{figure*}

The moment method relies on the MEC to reconstruct the flux distribution as a function of polar and azimuthal angle at all times.  To compare the evolved angular distributions between \flash and \emu, we show reconstructions of the flavor-diagonal density matrix distributions at a time $t=2\,{\rm ns}$ using mollwiede projections in Fig.\ \ref{fig:angular_distributions}.
The top row gives cell-averaged angular distributions for \flash with the MEC.
Columns 1-4 are for electron neutrino, $x$-flavor neutrino, electron antineutrino, and $x$-flavor antineutrino, respectively. We plot $\int d\epsilon\,\epsilon^2\varrho_{aa}$,
where $\varrho_{aa}$ is averaged over the simulation domain and the integral over energy $\epsilon$ serves to specify the units.
Column 5 is the ELN angular distribution $\int d\epsilon\,\epsilon^2[(\varrho_{ee}-\overline{\varrho}_{ee}) - (\varrho_{xx}-\overline{\varrho}_{xx})]$.
The bottom row gives the same flux quantities for \emu, where the average is done over cells and particles.
The granularity in the \emu plots is a result of the finite number of directions the particles can describe.

The mollwiede projections for the four angular distributions show excellent agreement at $t=2\, {\rm ns}$.
Notice also that the ELN remains less than or equal to zero, implying that the initial ELN crossing has been eradicated and that there is no longer a possibility of a FFO in both simulations.
However, the most visible difference between the two sets of calculations is indeed the ELN, as there is a larger range of ELN values in the \emu calculations.  This is the result of larger $\overline{\nu}_e$ fluences in the $\widehat{z}$ direction, i.e., a smaller amount of flavor transformation.

\section{Conclusions}

To summarize our results, we addressed the open question of whether or not a moment method would exhibit fast flavor transformation in neutron star mergers by designing a moment method appropriate for quantum neutrino transport, and applying it to classical neutron star merger conditions.
We found that fast flavor transformation did indeed occur although the moment method with MEC closure predicted more conversion of the electron neutrinos to other types but a similar level of mixing at the saturation peak.  Specifically, the final electron (on-diagonal) neutrino number density in the two-flavor moment method with MEC closure asymptoted to 64\% of its original value as compared with the two-flavor PIC calculation where it ended up at 74\% of its original value.  At the saturation peak, the moment method predicted  $|N_{ex}|$  to be  16\% of the original (on-diagonal) electron neutrino number density while the PIC calculation had 18\%. 
Additionally, when compared to PIC calculations as a baseline, the moment method captured the growth rate to 45 percent and the wavelength of the fastest growing mode in the neutrino field to 33 percent.  We attribute the differences between the moment method and the PIC method to our use of a classical closure which has previously been shown to overestimate the coherence of the moments \cite{Myers}. 

When including neutrino transport, \flash and other comparable state of the art codes take many millions of CPU hours for 3D calculations using classical moment based methods.  Not only does a quantum moment flavor transformation method offer the promise of more seamless integration, but also 
the computational time required to model the FFO physics is a factor of 30 less for the baseline \flash calculation as compared to \emu, which  
offers hope that a robust flavor transformation procedure can be coupled with a hydrodynamic code. 

For a moment method to be a viable option, we need to sample additional points that have a wide range of neutrino angular distributions in order to test the limits of the moment method \cite{usinprep}. In addition, future work will need to address several key issues.
For example, the bottom panel of Fig.\ \ref{fig:P_NSM} shows divergences between different resolutions during the latter decoherence phase. The centimeter length scales simulated here are also still orders of magnitude smaller than the kilometer scales simulated in state of the art merger simulations.
In addition, an investigation into quantum closures must be undertaken and the effect of differing closures \cite{2017MNRAS.469.1725M,Mezzacappa:2020oyq,PhysRevD.102.083017} on the results quantified.  
Finally, in order to capture the flavor physics in NSM or core collapse supernova scenarios, the momentum-changing collision term \cite{2019PhRvL.122i1101C,Richers2019,Johns:2021qby,Hansen:2022xza,2022arXiv220600676S} must be included. This is very challenging for computational-resource-heavy multi-angle neutrino schemes, but the moment method we have presented here is a step toward achieving this milestone.

\section*{Acknowledgments}

We acknowledge Julien Froustey, MacKenzie Warren and Don Willcox for useful conversations.  EG, JPK, and GCM are supported by the Department of Energy Office of Nuclear Physics award DE-FG02-02ER41216. EG acknowledges support by the National Science Foundation grant  No.\ PHY-1430152 (Joint Institute for Nuclear Astrophysics Center for the Evolution of the Elements). This work was partially enabled by the National Science Foundation under grant No.\ PHY-2020275 (Network for Neutrinos, Astrophysics, and Symmetries) and the Heising-Simons foundation under grant No.\ 2017-228. SR is supported by a NSF Astronomy and Astrophysics Postdoctoral Fellowship under award AST-2001760. FF acknowledges support from the Department of Energy through grant DE-SC0020435.
SMC is supported by the U.S. Department of Energy, Office of Science, Office of Nuclear Physics, Early Career Research Program under Award Number DE-SC0015904.
The calculations presented in this paper were undertaken on the \emph{Payne} machine at North Carolina State University which is supported in part by the Research Corporation for Science Advancement. This research used the Cori supercomputer of the National Energy Research Scientific Computing Center, a U.S. Department of Energy Office of Science User Facility located at Lawrence Berkeley National Laboratory, operated under Contract No.\ DEAC02-05CH11231.  This material is based upon work supported by the U.S. Department of Energy, Office of Science, Office of Advanced Scientific Computing Research and Office of Nuclear Physics, Scientific Discovery through Advanced Computing program under Award Number DE-SC0017955.
The authors would like to acknowledge the use of the following software: Matplotlib \cite{hunter:2007}, Numpy \cite{vanderwalt:2011}, and SciPy \cite{scipy}.

\bibliographystyle{elsarticle-num}
\bibliography{mff}

\end{document}